\documentclass[12pt]{iopart}

\usepackage{epsfig}
\def\lsim{\raise0.3ex\hbox{$<$\kern-0.75em\raise-1.1ex\hbox{$\sim$}}}
\def\gsim{\raise0.3ex\hbox{$>$\kern-0.75em\raise-1.1ex\hbox{$\sim$}}}
\newcommand{\hm}{\hat{m}}
\begin{document}

\hspace*{-0.5cm}
\mbox{} \hfill BNL-NT-07/7
\hspace*{-1.5cm}

\title{Transition temperature in QCD with physical light and strange
quark masses}

\author{Frithjof Karsch (for the RBC-Bielefeld collaboration)}

\address{Physics Department, Brookhaven National Laboratory,
Upton, NY 11973, USA}

\ead{karsch@bnl.gov}

\begin{abstract}
We present results from a calculation of the transition temperature in QCD with
two light (up, down) and one heavier (strange) quark mass as well as for QCD 
with three degenerate quark masses. Furthermore, we discuss first results 
from an ongoing calculation of the QCD equation of state with almost 
realistic light and strange quark masses.
\end{abstract} 

\section{Introduction}

Lattice calculations provide a unique first principle approach to
the thermodynamics of QCD. They have the potential to produce definitive
answers for basic bulk thermodynamic properties as well as the phase 
diagram of QCD. To reach this ambitious goal 
one has to control lattice discretization errors
that lead to systematic deviations of lattice calculations from continuum
physics, as well as finite volume effects that that can obscure the 
thermodynamic limit. 

We present here results from a detailed analysis of the transition temperature
in QCD with an almost physical value of the strange quark mass and several
values of the light quark mass. This allows to perform an extrapolation to 
the chiral limit as well as to the physical point. Calculations have been 
performed for two different values of the lattice cut-off which allows to 
estimate
the lattice discretization errors and perform an extrapolation to the continuum
limit. The flavor dependence of these results are examined through an analogous
analysis of QCD with three degenerate quark flavors. 
We also report on results from an ongoing study of the QCD equation of
state with almost physical light and strange quark masses. 
All these calculations have been performed with an ${\cal O}(a^2)$ tree-level
improved gauge action and an improved staggered fermion action (p4fat3) that 
also improves bulk thermodynamic observables  at ${\cal O}(a^2)$. For 
further details on the calculational set-up we refer to \cite{rbcBI2+1,rbcBI3}.

\section{The transition temperature}

\begin{figure}[t]
\begin{center}
\epsfig{file=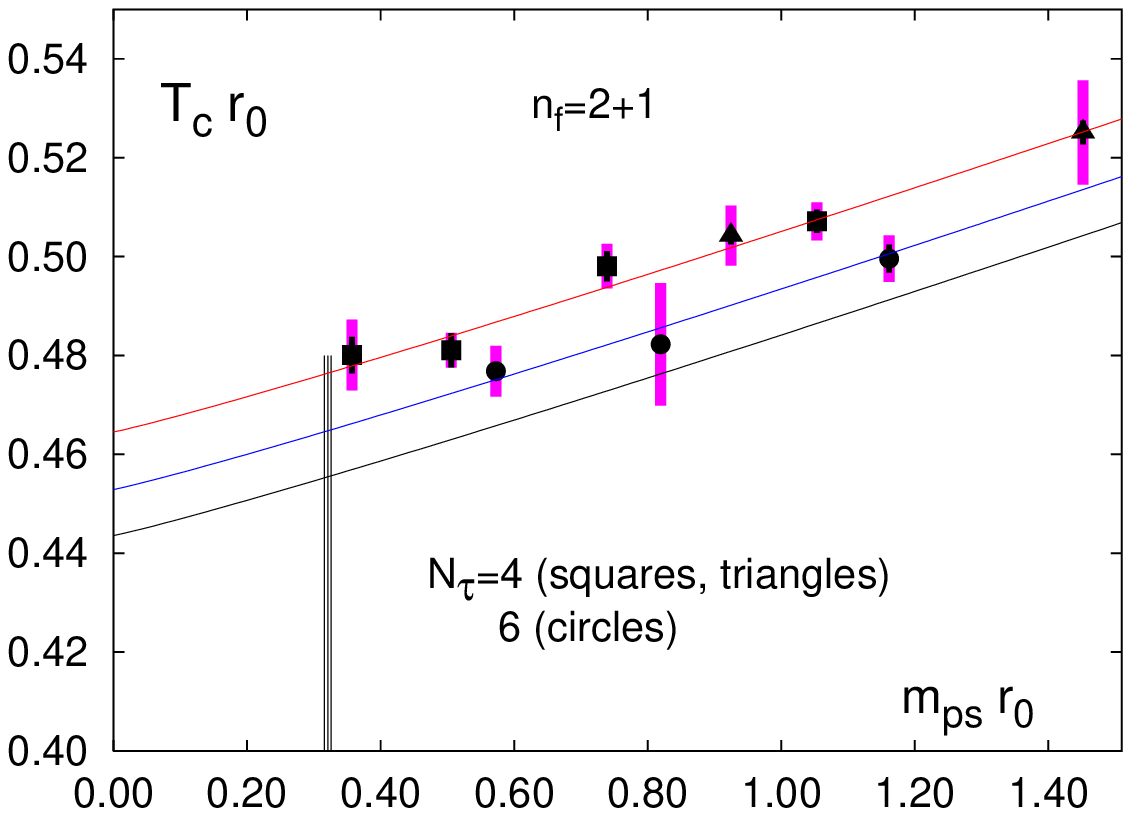,width=74mm}
\epsfig{file= 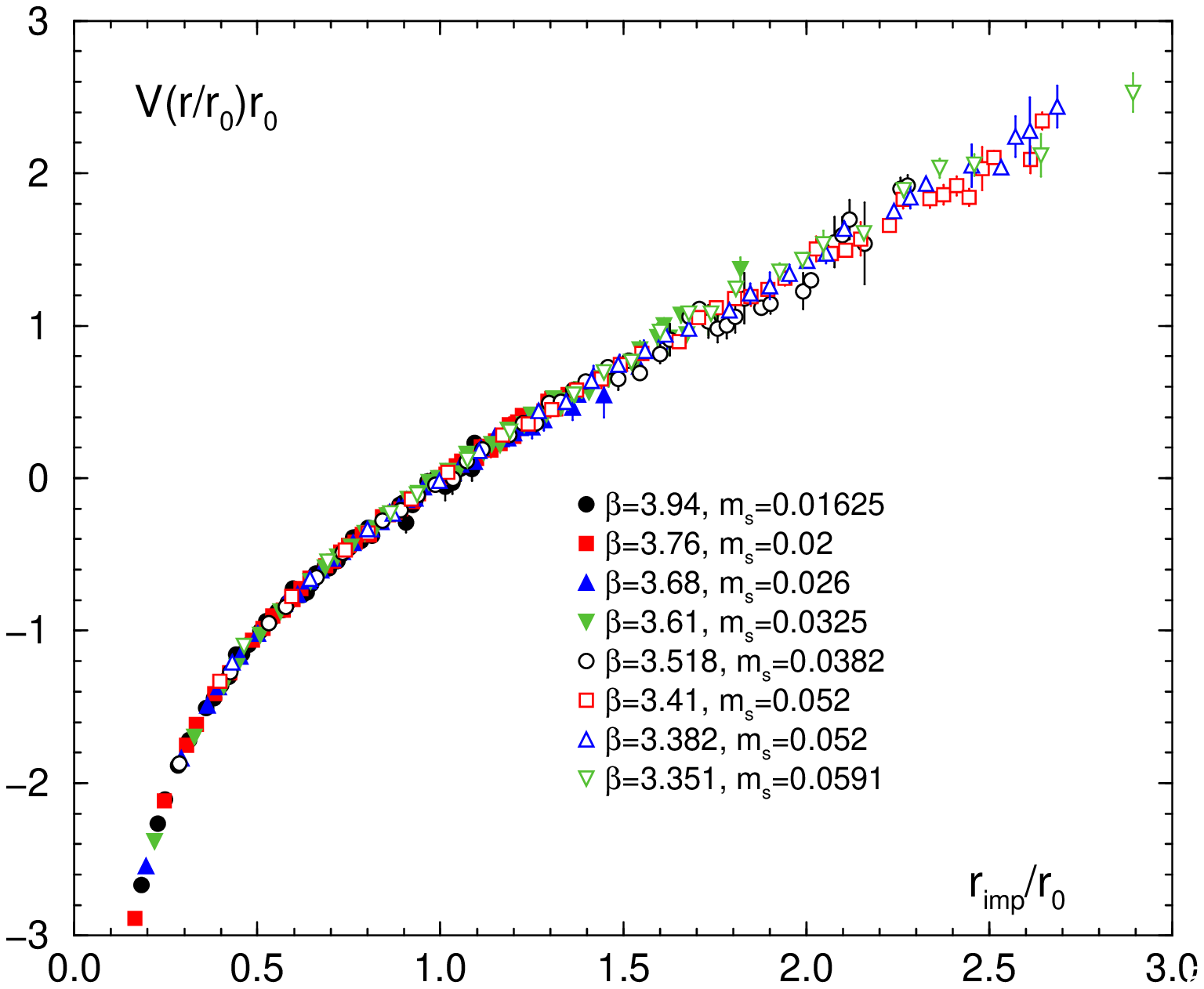,width=74mm}
\end{center}
\caption{\label{fig:chiral} The transition temperature (left) calculated
on lattices with temporal extent $N_\tau =4,~6$ and the static quark
potential (right) in units of $r_0$ calculated 
on $16^3\times 32$ lattices. $V(r)$ has been calculated for a wide range 
of gauge coupling $\beta$ in which the lattice cut-off varies by a factor $4$.
This range also covers the the relevant parameter range 
of the finite temperature calculations.
}
\end{figure}

In order to determine the transition temperature and its dependence on 
the light quark mass we have performed calculations with a fixed 
value of the bare strange quark mass and several (degenerate) values for 
the two light quark masses that correspond to light pseudo-scalar
masses ($m_{ps}$) ranging from $m_{ps}\simeq 150$~MeV to 
$m_{ps}\simeq 500$~MeV. For non-zero values of the quark masses the
transition to the high temperature phase of QCD is signaled by peaks in
response functions. We have analyzed  Polyakov loop and chiral  
susceptibilities to locate
the transition point for the different quark mass values on lattices with
temporal extent $N_\tau =4$ and $6$ \cite{rbcBI2+1}. 
The spatial extent of the lattice has
been varied between $N_\sigma/N_\tau = 2$ and $4$. We generally observe
only a weak volume dependence of the transition point and also find that
the above susceptibilities as well as quark number susceptibilities in the 
light and strange quark sector \cite{schmidt} yield, 
within statistical errors, consistent results for the transition point. 

In order to convert the results for the transition temperature to 
physical values one needs to set the scale through the calculation of
zero temperature observables. Unlike hadronic observables in the light
quark sector, which require a careful handling of chiral extrapolations,
are known to be difficult to determine and in some cases are influenced by 
large discretization errors, the parameters of 
the static quark potential have been found to show only a mild dependence
of the lattice cut-off \cite{milcr0}. We thus have performed extensive 
studies of the static quark potential, $V(r)$, on zero 
temperature lattices and extracted the scale parameter $r_0$ defined 
through the slope of $V(r)$, {\it i.e.} 
$\displaystyle{\left(r^2 {\rm d}V(r)/{\rm d}r\right)_{r=r_0} = 1.65}$.
Using the results obtained for $r_0$ in lattice units at the crossover 
couplings, which have been determined  
in finite temperature simulations on lattices with temporal extent 
$N_\tau = 4$ and $6$, the transition temperatures can then be calculated as 
$T_cr_0 \equiv r_0/N_\tau a$.  
Results for $T_cr_0$ and the static quark potential
in units of $r_0$ are shown in Fig.~\ref{fig:chiral}. Results for $T_cr_0$
have then been extrapolated to the continuum limit using an ansatz that
incorporates the quark mass dependence of $T_c$ through the calculated
values of the lightest pseudo-scalar mass as well as a quadratic
dependence on the lattice cut-off,
\begin{equation}
(T_c r_0)_{ \hm_l,\hm_s,N_\tau } = (T_c r_0)_{ 0,m_s,\infty } +
A (m_{ps}r_0)^d + B/N_\tau^2  \; .
\label{extrapolation}
\end{equation}
Here $d$ has been varied between $d=1$ and $2$ to allow for uncertainties
in the extrapolation to the chiral limit \cite{rbcBI2+1}. The continuum 
extrapolation is
shown as the lower curve in Fig.~\ref{fig:chiral}(left). 
For the transition temperature at the physical point we find,
\begin{equation}
T_c r_0  = 0.457(7)^{+8}_{-2} \;\; \Leftrightarrow \;\;
T_c = 192(7)(4)\; {\rm MeV}\;\; ,
\end{equation}
where we have used $r_0 = 0.469(7)$~fm  
\cite{gray} 
to set the scale for $T_c$. 

A similar analysis has been performed for QCD with three degenerate quark 
masses \cite{rbcBI3}.
In this case the transition becomes first order for small but non-zero
values of the quark mass. We find for the transition temperature in the chiral 
limit
$T_c r_0= 0.419(9)$ while in the same limit the transition temperature in 
(2+1)-flavor QCD  is
$T_c r_0= 0.444(6)(12)$. We note that $T_c r_0$ in 3-flavor
QCD is about 5\% smaller than in (2+1)-flavor QCD. Moreover, the difference
between the transition temperature in the chiral limit and at the physical 
point is only about 3\%. This is a direct consequence of the weak quark mass
dependence of the transition temperature, which has been observed earlier
\cite{peikertTc}. A similarly weak quark mass dependence has also been
observed in studies performed with another version of an improved staggered 
fermion action (asqtad) on lattices with temporal extent $N_\tau = 4,~6$ and
$8$ \cite{milc}. This analysis, however, lead to a 10\% smaller transition 
temperature, $T_c r_0 = 0.402 (29)$ than the value reported here. 
To resolve
the origin of this discrepancy needs further investigations.

\section{The QCD equation of state}

\begin{figure}
  \includegraphics[height=.24\textheight]{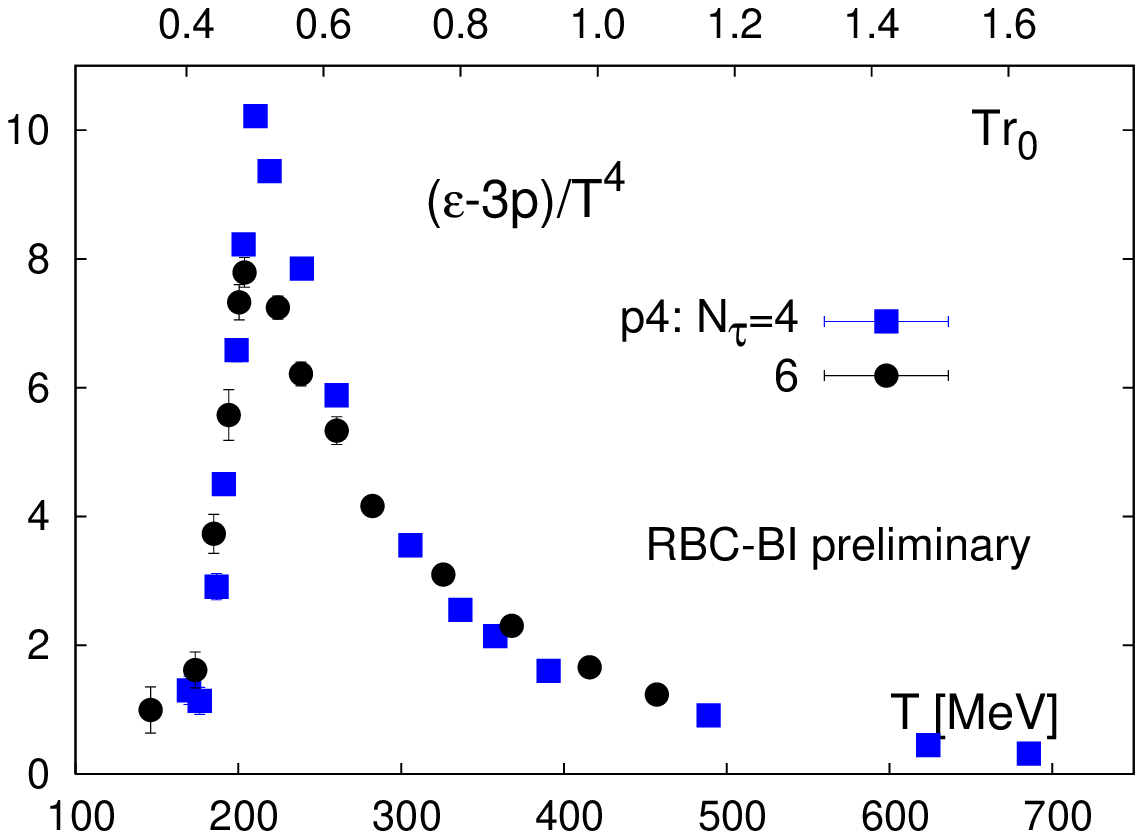}
  \includegraphics[height=.23\textheight]{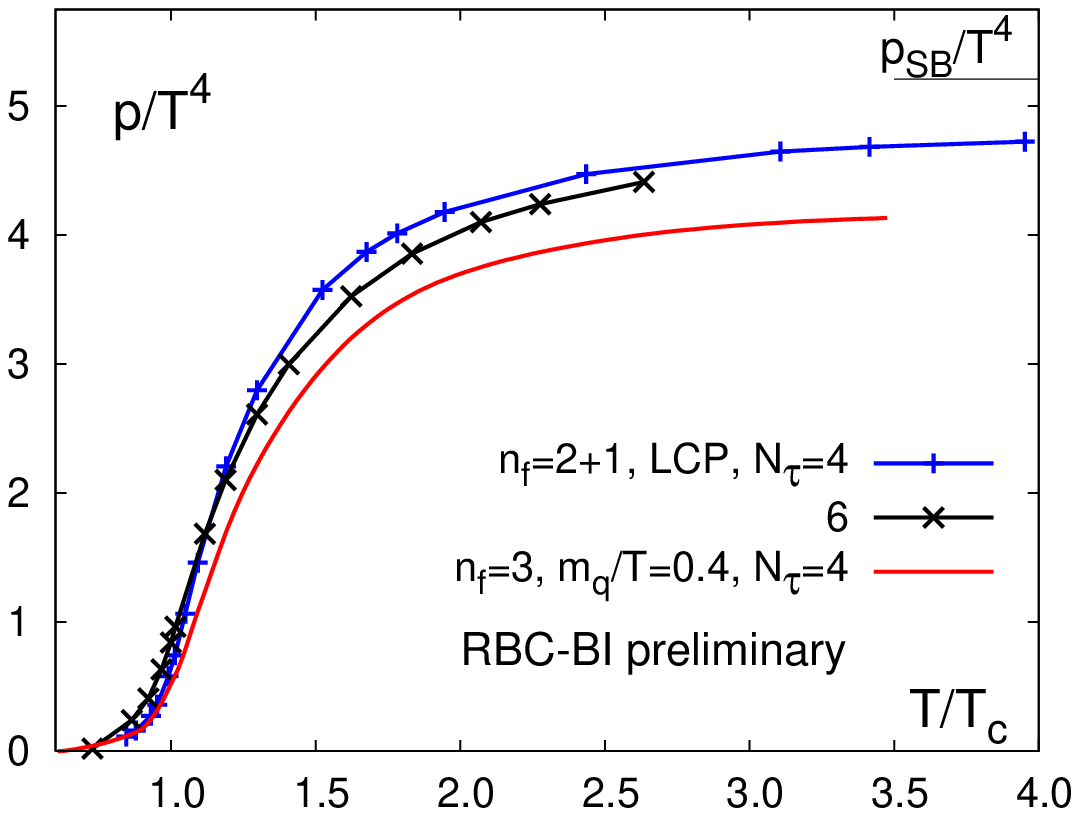}
  \caption{The trace anomaly, $(\epsilon -3p)/T^4$, calculated in (2+1)-flavor 
QCD on lattices with temporal extent $N_\tau =4$ and $6$, respectively. 
Fig.2(right) shows the resulting pressure from Eq.~\ref{pres} as well as 
earlier results obtained for 3-flavor QCD (see text).
}
\label{fig:eos}
\end{figure}

We currently study the equation of state in (2+1)-flavor QCD \cite{rbcBIeos}
on lattices with temporal extent $N_\tau =4$ and $6$ using the
same calculational set-up used for the study of the transition temperature.
Calculations are performed along a line of constant physics
characterized by a fixed strange pseudo-scalar mass $m_{\bar{s}s} \simeq
660$~MeV and a fixed ratio of $m_{\bar{s}s}$ and the kaon mass, 
$m_{\bar{s}s}/m_K \simeq 1.3$. The light pseudo-scalar (pion) mass on 
this trajectory is about $220$~MeV.

In Fig.~\ref{fig:eos} we show results for the trace anomaly, 
$\Theta^{\mu}_\mu (T)\equiv \epsilon -3p$ 
in units of $T^4$ and the pressure ($p$) which is 
deduced from $ \Theta^{\mu}_\mu (T)$ using 
standard thermodynamic relations, 
\begin{equation}
\frac{p(T)}{T^4} - \frac{p(T_0)}{T_0^4} = \int_{T_0}^{T} {\rm d}T'
\frac{1}{T'^5} \Theta^{\mu}_\mu (T') \;\; .
\label{pres}
\end{equation}
We also included in Fig.~\ref{fig:eos}(right) results for the pressure 
obtained earlier \cite{peikert}
for 3-flavor QCD with larger quark masses that have been held fixed in 
units
of $T$. For these earlier calculations the light pseudo-scalar mass
was about $770$~MeV in the transition region and would have been about
$2$~GeV in the region around $T\simeq 3T_c$! 

The results presented here for $(\epsilon-3p)/T^4$ and $p/T^4$ in 
(2+1)-flavor QCD are entirely consistent with results obtained with the
asqtad action \cite{milceos}. In particular, results for $p/T^4$ 
obtained with the p4fat3 and asqtad actions, which both have a ${\cal O}(a^2)$ 
improved ideal gas limit, show little
cut-off dependence in the entire temperature regime analyzed. This is
in contrast to calculations performed with a variant of standard staggered 
fermions \cite{fodoreos} which unlike p4fat3 and asqtad actions lead to 
large cut-off distortions of the 
high-T limit on lattices with temporal extent $N_\tau =4$ and $6$. 
We also note that we find no evidence for 
a significant broadening of the transition region when going closer to
the continuum limit. Whether this happens for larger $N_\tau$, as a recent
analysis of the transition temperature on lattices with temporal extent up to
$N_\tau=10$ may suggest \cite{fodorTc}, has to be examined in more detail.

\ack
%\section*{Acknowledgments}
This manuscript has been authored under contract number
DE-AC02-98CH1-886 with the U.S. Department of Energy.
%Accordingly,
%the U.S. Government retains a non-exclusive, royalty-free license to
%publish or reproduce the published form of this contribution, or allow
%others to do so, for U.S.~Government purposes.

\end{document}